\documentclass[sigconf]{acmart}

\usepackage{mychangebar}
\usepackage{booktabs} 
\usepackage{listings}
\usepackage{algorithm2e}
\usepackage{subcaption}
\usepackage{highlighter}
\usepackage{todonotes}
\usepackage{enumitem}
\setlist{leftmargin=4mm}

\usepackage[nohyphen]{underscore}
\usepackage[english]{babel}

\newcommand{\xqed}[1]{%
 \leavevmode\unskip\penalty9999 \hbox{}\nobreak\hfill
  \quad\hbox{\ensuremath{#1}}}

\newcommand*{\qef}{\xqed{\scriptstyle{\blacksquare}}}

\newcommand*{\tinyqed}{\xqed{\scriptstyle{\square}}}

\newcommand{\eqdef}{\buildrel \mbox{\tiny\textrm{def}} \over =}
\newcommand{\errorimplies}{\overset{e}{\Rightarrow}}
\newcommand{\appliesto}{\triangleright}
\newcommand{\satisfiedby}{\blacktriangleright}

\DeclareMathOperator*{\confidence}{conf}
\DeclareMathOperator*{\support}{supp}
\DeclareMathOperator*{\maximize}{maximize}

\renewcommand\footnotetextcopyrightpermission[1]{} 

\setcopyright{none}

\begin{document}

\lstset{
    language=C,
    basicstyle=\footnotesize\ttfamily,
    breaklines=true,
    numberstyle=\scriptsize,
    numbers=left,
    frame=none,
    xleftmargin=2em,
    framexleftmargin=2em,
    tabsize=2,
    escapeinside={/*@}{@*/},
    keywordstyle=\color{blue},
    commentstyle=\color{gray}
}

\newcommand{\code}[1]{\texttt{#1}}

\newcommand{\hl}[1]{\colorbox{lightgray}{#1}}
\newcommand{\Omit}[1]{}

\crefname{section}{\S\!}{\S\S}
\crefname{table}{Table}{Tables}
\crefname{figure}{Figure}{Figures}
\crefname{subfigure}{Figure}{Figures}
\crefname{definition}{Definition}{Definitions}
\crefname{equation}{Equation}{Equations}
\crefname{example}{Ex.}{Examples}
\crefname{algorithm}{Algorithm}{Algorithms}

\newcommand{\programtovec}[0]{\texttt{func2vec}\xspace}
\newcommand{\lpds}[0]{$\ell$-PDS\xspace}

\newcommand{\fscore}[0]{0.77}
\newcommand{\avgprecision}[0]{87\%}
\newcommand{\avgrecall}[0]{71\%}
\newcommand{\goldensetfunctions}[0]{683}
\newcommand{\goldensetrelations}[0]{9,600}
\newcommand{\goldensetmust}[0]{7,822}
\newcommand{\goldensetmustnot}[0]{1,778}
\newcommand{\goldensetclasses}[0]{127}

\newcommand{\todostructure}[1]{\todo[inline,color=yellow!40]{STRUCTURE: #1}}
\newcommand{\todolater}[1]{\todo[inline,color=green!40]{LATER: #1}}
\newcommand{\todonow}[1]{\todo[inline,caption={}]{TODO: #1}}
\newcommand{\tododelete}[1]{\todo[inline,color=red!40]{DELETE: #1}}

\newcommand{\todocam}[1]{\todo[inline]{CAMERA: #1}}
\newcommand{\todoresub}[1]{\todo[inline]{RESUBMIT: #1}}
\newcommand{\todowhen}[1]{\todo[inline]{WHEN?: #1}}

\newcommand{\intrarule}[2]{\langle \mathrm{p, #1} \rangle \hookrightarrow
  \langle \mathrm{p, #2} \rangle}
\newcommand{\callrule}[3]{\langle \mathrm{p, #1} \rangle \hookrightarrow
  \langle \mathrm{p, #2\ #3} \rangle}
\newcommand{\returnrule}[1]{\langle \mathrm{p, #1} \rangle \hookrightarrow
  \langle \mathrm{p}, \epsilon \rangle}

\newcommand{\functionlabel}[1]{\textbf{\texttt{#1}}}
\newcommand{\typelabel}[1]{\emph{#1}}
\newcommand{\otherlabel}[1]{\MakeUppercase{#1}}
\newcommand{\errorlabel}[1]{\MakeUppercase{\texttt{#1}}}

\SetKwFunction{RandomKw}{Random}
\SetKwFunction{TrainModelKw}{TrainModel}

\newcommand{\subsubsubsection}[1]{\smallskip\noindent\textbf{#1}}

\makeatletter
\newcommand{\xRightarrow}[2][]{\ext@arrow 0359\Rightarrowfill@{#1}{#2}}
\makeatother

\newcommand{\deleted}[1]{
  \cbcolor{red}
  \begin{changebar}
    #1
  \end{changebar}
  }

\newcommand{\added}[1]{
  \cbcolor{green}
  \begin{changebar}
    #1
  \end{changebar}
  }

\newcommand{\changed}[1]{
  \cbcolor{gray}
  \begin{changebar}
    #1
  \end{changebar}
  }

\title{Path-Based Function Embedding \\ and its Application to
  Specification Mining}

\author{Daniel DeFreez, Aditya V. Thakur, Cindy Rubio-Gonz\'alez}
\affiliation{%
  \institution{University of California, Davis}
}
\email{{dcdefreez, avthakur, crubio}@ucdavis.edu}

\begin{abstract}
Identifying the relationships among program elements is useful for program
understanding, debugging, and analysis. One such relationship is synonymy.
Function synonyms are functions that play a similar role in code, e.g. functions
that perform initialization for different device drivers, or functions that
implement different symmetric-key encryption schemes. Function synonyms are not
necessarily semantically equivalent and can be syntactically dissimilar;
consequently, approaches for identifying code clones or functional equivalence
cannot be used to identify them. This paper presents \programtovec, an algorithm
that maps each function to a vector in a vector space such that function
synonyms are grouped together. We compute the function embedding by training a
neural network on sentences generated from random walks over an encoding of the
program as a labeled pushdown system (\lpds). We demonstrate that 
\programtovec is effective at identifying function synonyms in the Linux kernel. Furthermore,
we show how function synonyms enable mining error-handling
specifications with high support in Linux file systems and drivers.
\end{abstract}

\settopmatter{printfolios=true} 

\maketitle

\vspace{-1ex}

\section{Introduction}
\label{sec:introduction}

Apart from writing new code, a software engineer spends a substantial
amount of time understanding, evolving, and verifying existing
software.  \emph{Program
  comprehension}~\citep{Maalej:2014:CPC:2668018.2622669} entails
inferring a mental model of the relationships among various program
elements~\cite{letovsky1987cognitive}. When available, documentation
can aid program comprehension~\cite{kajko2005survey}.  For instance,
documentation about high-level API functions often contains a ``See
Also'' section listing other related functions, enabling the reader to
navigate to a different, relevant portion of the API.  However, such
documentation is almost never available for low-level code such as the
Linux kernel. Even if such documentation is available, it is difficult
to keep it up to date as the code evolves
\cite{lethbridge2003software}. Furthermore, Linux is written in C, a
language lacking features such as polymorphism and encapsulation that
make explicit the relationships between functions.

Identifying relationships among functions is challenging because
related functions are often semantically different and syntactically
dissimilar. For example, the functions \code{snd_atiixp_free} and
\code{snd_intel8x0_free} in the Linux device drivers \code{atiixp} and
\code{intel8x0}, respectively, are semantically different, but serve
the same purpose in these device drivers. We refer to such functions
as \emph{function synonyms}. The above functions follow a naming
convention, but that is not necessarily the case. For example,
\texttt{acpi_video_get_brightness} and
\texttt{intel_panel_get_backlight} each return the brightness level of
the backlight.  Conversely, functions with similar names are not
necessarily synonyms. Consider \texttt{rcu_seq_start} which adjusts
the current sequence number, and \texttt{kprobe_seq_start} which
merely returns the current sequence number. Because of the semantic
and syntactic differences in the code, and because naming conventions
are not a reliable indicator of similarity, techniques that identify
code clones, check semantic equivalence, or rely on naming conventions
cannot be used.

\begin{figure}[!t]
    \centering
    \vspace{2em}
    \includegraphics[scale=0.5]{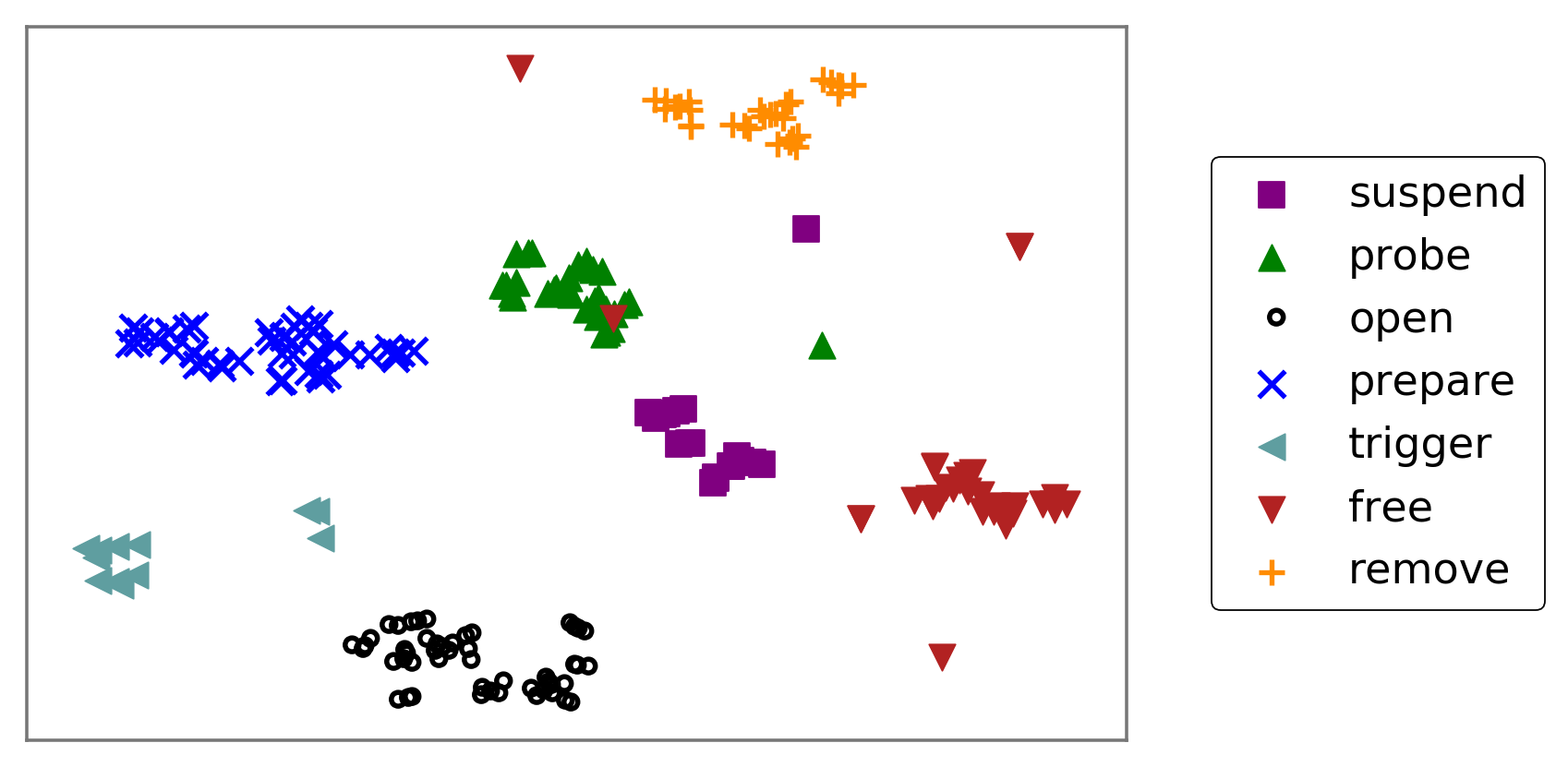}
    \caption{Function Synonym Clusters}
    \label{fig:tsne-intro}
    \vspace{-4ex}
\end{figure}

This paper presents \programtovec, a technique that computes a map
from each function to a vector in a continuous vector space such that
vectors for function synonyms are in close proximity \emph{without}
any previous knowledge about naming conventions. 
\Cref{fig:tsne-intro} illustrates the output of \programtovec for a
subset of functions in Linux; in particular, \programtovec maps each
function to a vector in $\mathbb{R}^{300}$, which is then projected
onto 2-dimensions using t-SNE~\citep{maaten2008visualizing}. Functions
that play the same role in different components --- function synonyms
--- are close together in the \programtovec embedding, forming
clusters. \Cref{fig:tsne-intro} shows such clusters in the PCI sound
drivers. Function synonyms are grouped by functionality (\emph{probe},
\emph{open}, \emph{prepare}, \emph{free}, etc.). For example, the
functions \code{snd_atiixp_free} and \code{snd_intel8x0_free} both
belong to the cluster labeled \emph{free}.

\emph{This technique is the first to use static program traces to
  learn a function embedding that captures the hierarchical structure
  of programs.}  Specifically, we encode the program as a labeled
pushdown system (\lpds), where labels are used to represent various
program elements such as function calls, instructions, types, and
error codes. We generate random walks over the \lpds, and use these
walks to learn a vector embedding using a neural
network~\citep{word2vec}.

To demonstrate the effectiveness of \programtovec, we create a
distributed representation of functions for a runnable Linux kernel (2
million LOC). \emph{This paper is the first to apply such a technique
  to large-scale low-level code such as the Linux kernel}. We evaluate
\programtovec on a manually created gold standard of
\goldensetfunctions\ Linux file system and driver functions grouped
into \goldensetclasses\ classes. Our evaluation shows that
\programtovec is capable of identifying relationships between
functions with both high precision and recall, \avgprecision\ and
\avgrecall\ respectively.

Furthermore, we present a case study showing how function synonyms
identified by \programtovec can be used to find software
specifications from code.  Specification mining is a popular technique
with many applications \citep{DBLP:conf/fase/AcharyaX09,
  DBLP:conf/icse/ThummalapentaX09, DBLP:conf/tacas/GouesW09,
  DBLP:conf/dsn/SahaL0LM13}. In general, specification mining requires
a large number of supporting examples for each specification to avoid
false positives, but even for a code base as large as Linux, many
specifications do not have enough examples to reach a useful support
(number of occurrences) threshold
\citep{DBLP:conf/dsn/SahaL0LM13}. This problem is further exacerbated
for \emph{error-handling} specifications; error-handling code is not
as common as normal-execution code.

In this paper, we leverage \programtovec to use multiple
implementations of Linux file systems and drivers to obtain
\emph{cross-implementation} error-handling specifications that have
high support. A cross-implementation specification is one that
consists of multiple error-handling specifications that would be the
same if we could identify certain functions to be synonyms. An example
of such a specification is described in
\cref{sec:motivating-example}. To show the usefulness of \programtovec
in this context, we devise and implement an algorithm for inferring
error-handling specifications. 
We evaluate the effect of using \programtovec results on
mining for 5 Linux file systems, and 48 Linux device drivers, and show
that using \programtovec indeed yields better specifications.

The contributions of this paper are as follows:
\begin{itemize}
\item \programtovec, a technique for learning a function embedding
  that captures the hierarchical structure of programs
  (\cref{sec:approach}).
\item An evaluation of the effectiveness of \programtovec for finding
  function synonyms in the Linux kernel
  (\cref{sec:programtovec-eval}).
\item A formulation of error-handling specifications for low-level systems
  code (\cref{sec:ehnfer}).
\item An evaluation of the usefulness of \programtovec for mining
  cross-implementation error-handling specifications across 5 Linux
  file systems, and 48 Linux device drivers (\cref{sec:ehnfer-eval}).
\end{itemize}

\noindent We describe related work in \cref{sec:related}, and conclude in
\cref{sec:conclusions}.

\section{Motivating Example}
\label{sec:motivating-example}

\begin{figure*}
\begin{subfigure}[b]{.5\textwidth}
  \tiny
\begin{lstlisting}[xleftmargin=1em,numbersep=1pt,firstnumber=1585,linerange={1-1}]
int snd_atiixp_create(struct snd_card *card){
\end{lstlisting}
\begin{lstlisting}[xleftmargin=1em,numbersep=1pt,firstnumber=1595,linerange={3-10}]
 if ((err=/*@\bh@*/pci_enable_device(pci)/*@\eh@*/) < 0) /*@\color{red}{//H1} @*/ /*@\label{atiixp:enable}@*/
   return err; /*@\label{atiixp:eh1}@*/

 chip = kzalloc(sizeof(*chip), ...); /*@\label{atiixp:kzalloc}@*/
 if (chip == NULL) { /*@\color{red}{//H2} @*/ /*@\label{atiixp:ifkzalloc}@*/
   /*@\fbox{pci_disable_device(pci);}@*/ /*@\label{atiixp:eh2-begin}@*/
   return -ENOMEM; /*@\label{atiixp:eh2-end}@*/
  }
\end{lstlisting}
\begin{lstlisting}[xleftmargin=1em,numbersep=1pt,firstnumber=1609,linerange={12-17}]
 if((err=/*@\bh@*/pci_request_regions(pci)/*@\eh@*/) < 0){/*@\color{red}{ //H3} @*/ /*@\label{atiixp:regions}@*/
   pci_disable_device(pci); /*@\label{atiixp:eh3-begin}@*/
   kfree(chip);
   return err; /*@\label{atiixp:eh3-end}@*/
 }
 chip->addr = pci_resource_start(pci, 0);
\end{lstlisting}
\begin{lstlisting}[xleftmargin=1em,numbersep=1pt,firstnumber=1622,linerange={19-23}]
 if (request_irq(pci->irq,...)) { /*@\color{red}{//H4} @*/ /*@\label{atiixp:irq}@*/
   dev_err(card->dev, "IRQ %d", pci->irq); /*@\label{atiixp:eh4-begin}@*/
   /*@\fbox{snd_atiixp_free(chip);}@*/ /*@\label{atiixp:free}@*/
   return -EBUSY; /*@\label{atiixp:eh4-end}@*/
 }
\end{lstlisting}
\begin{lstlisting}[xleftmargin=1em,numbersep=1pt,firstnumber=1632,linerange={25-28}]
 if((err=snd_device_new(card,...) < 0)) {/*@\color{red}{ //H5} @*/ /*@\label{atiixp:device-new}@*/
   /*@\fbox{snd_atiixp_free(chip);}@*/  /*@\label{atiixp:eh5-begin}@*/
   return err; /*@\label{atiixp:eh5-end}@*/
 }
\end{lstlisting}
\begin{lstlisting}[xleftmargin=1em,numbersep=1pt,firstnumber=1639,linerange={30-30}]
}
\end{lstlisting}
  \vspace{-2ex}  
  \caption{Function \code{snd_atiixp_create} in driver atiixp.}
  \label{fig:spec-atiixp}
\end{subfigure}
\begin{subfigure}[b]{.49\textwidth}
  \tiny
\begin{lstlisting}[xleftmargin=1em,numbersep=1pt,firstnumber=2989,linerange={1-1}]
int snd_intel8x0_create(struct snd_card *card){
\end{lstlisting}
\begin{lstlisting}[xleftmargin=1em,numbersep=1pt,firstnumber=3036,linerange={3-10}]
 if ((err=/*@\bh@*/pci_enable_device(pci)/*@\eh@*/) < 0) /*@\color{red}{//H1} @*/ /*@\label{intel8x0:enable}@*/
   return err; /*@\label{intel8x0:eh1}@*/

 chip = kzalloc(sizeof(*chip), ...);/*@\label{intel8x0:kzalloc}@*/
 if (chip == NULL) {/*@\color{red}{ //H2} @*/ /*@\label{intel8x0:intel8x0-null}@*/
   /*@\fbox{pci_disable_device(pci);}@*/ /*@\label{intel8x0:eh2-begin}@*/
   return -ENOMEM; /*@\label{intel8x0:eh2-end}@*/
 }
\end{lstlisting} 
\begin{lstlisting}[xleftmargin=1em,numbersep=1pt,firstnumber=3062,linerange={12-16}]
 if ((err=/*@\bh@*/pci_request_regions(pci)/*@\eh@*/) < 0){/*@\color{red}{ //H3} @*/ /*@\label{intel8x0:regions}@*/
   kfree(chip); /*@\label{intel8x0:eh3-begin}@*/
   pci_disable_device(pci);
   return err; /*@\label{intel8x0:eh3-end}@*/
 }
\end{lstlisting}
\begin{lstlisting}[xleftmargin=1em,numbersep=1pt,firstnumber=3179,linerange={18-27}]
 if ((err = snd_intel8x0_chip_init()) < 0) {/*@\color{red}{ //H4} @*/ /*@\label{intel8x0:init}@*/
   /*@\fbox{snd_intel8x0_free(chip);}@*/  /*@\label{intel8x0:free1}@*/
   return err;
 }
 
 if (request_irq(pci->irq, ...)) { /*@\color{red}{//H5} @*/ /*@\label{intel8x0:irq}@*/
   dev_err(card->dev, "IRQ %d", pci->irq); /*@\label{intel8x0:eh4-begin}@*/
   /*@\fbox{snd_intel8x0_free(chip);}@*/  /*@\label{intel8x0:free2}@*/
   return -EBUSY; /*@\label{intel8x0:eh4-end}@*/
 }
\end{lstlisting}
\begin{lstlisting}[xleftmargin=1em,numbersep=1pt,firstnumber=3200,linerange={29-29}]
}
\end{lstlisting}
  \vspace{-2ex}    
  \caption{Function \code{snd_intel8x0_create} in driver intel8x0.}
  \label{fig:spec-intel8x0}
\end{subfigure}
\caption{(a) An excerpt from the function \texttt{snd_atiixp_create} in
  the atiixp driver (sound/pci/atiixp.c). The function contains
  \emph{two} error-handling specifications. Each specification
  consists of a context set (function calls highlighted in gray) and a
  response set (function calls in a box). The first specification is
  associated with error handler \texttt{H2} and has a 1-element
  context (highlighted in gray) and a 1-element response (in a
  box). The second specification is associated with handlers
  \texttt{H4} and \texttt{H5}. It has a 2-element context (highlighted
  in gray) and a 1-element context (in a box). Fig. 1. (b) An excerpt
  from the function \texttt{snd_intel8x0_create} in the intel8x0
  driver (sound/pci/intel8x0.c) in which \emph{two} similar
  error-handling specifications are found. The specifications across
  the two drivers are similar except for the functions
  \texttt{snd_atiixp_free}, and \texttt{intel8x0_free}. Since we
  identify these functions as synonyms, we refer to the set
  of corresponding specifications as cross-implementation
  specifications.}
\label{fig:all}
\vspace{-2ex}
\end{figure*}


In this section, we present a real-world example in which identifying
function synonyms can be useful. Specifically, we describe an
error-handling specification found across various Linux PCI sound
drivers.  An error handler is a piece of code that is executed if an
error occurs. For Linux code (written in C), each error handler
corresponds to a conditional statement that checks for an error. An
error-handling specification imposes requirements on an error handler.

\cref{fig:spec-atiixp} shows an excerpt from the function
\texttt{snd_atiixp_create} in the atiixp sound driver. In particular,
it shows five error handlers (marked \texttt{H1} through
\texttt{H5}). Here we describe a specification associated with the
error handler \texttt{H4} on \cref{atiixp:irq}. The specification is
expressed as \{\texttt{pci_enable_device, pci_request_regions}\}
$\errorimplies$ \{\texttt{snd_atiixp_free}\} (simplified for clarity
of explanation), which says that whenever functions
\texttt{pci_enable_device} and \texttt{pci_request_regions} are
successfully called, and an unrelated error occurs later on, then the
function \texttt{snd_atiixp_free} must be called to release the
resources acquired by \texttt{pci_enable_device} and
\texttt{pci_request_regions}. In the specification, the actions in the
set before the arrow correspond to the \emph{context}, and the actions
in the set after the arrow refer to the \emph{response}; see
\cref{sec:ehnfer}. This is an \emph{error-handling} specification
because it applies if an error occurs in the given context. In the
figure, we highlight the context in gray, and place the response
actions in a box. The support of the specification is 2; that is, we
find only 2 occurrences that follow this specification.
Error-handling code is not as common as normal-execution code, thus
error-handling specifications often have low
support~\citep{DBLP:conf/dsn/SahaL0LM13}.

\cref{fig:spec-intel8x0} shows a specification found in the intel8x0
driver. Note that the code fragments in \cref{fig:all} look almost
identical because we have not shown the irrelevant code (36 LOC in
\texttt{snd_atiixp_create} and 193 LOC in
\texttt{snd_intel8x0_create}). The intel8x0 specification is expressed
as \{\texttt{pci_enable_device, pci_request_regions}\} $\errorimplies$
\{\texttt{snd_intel8x0_free}\}, which is the same as the atiixp
specification except for the response. This specification has a
support of 7.  However, \programtovec reports that
\texttt{snd_atiixp_free} and \texttt{snd_intel8x0_free} are function
synonyms, thus the above specifications describe a
\emph{cross-implementation} error-handling specification. Note that
the above specification is also associated with handler \code{H5} as
it shares the same context, and has the same response as \code{H4}.

The advantage of finding cross-implementation specifications is that
their support is higher than those of the corresponding individual
specifications. For example, when considering 48 device drivers,
\programtovec finds 12 additional function synonyms for
\texttt{snd_atiixp_free} and \texttt{snd_intel8x0_free}. Using this
information, we find a cross-implementation specification with a
support of 57. As can be seen, function synonyms are crucial for
finding cross-implementation specifications with high support. The
rest of the paper describes how \programtovec creates function
embeddings of programs to find function synonyms, and how this
information can be used to enable cross-implementation specification
mining.

\section{Func2Vec: Path-Based Function Embedding}
\label{sec:approach}

\begin{figure}
\includegraphics[scale=0.8]{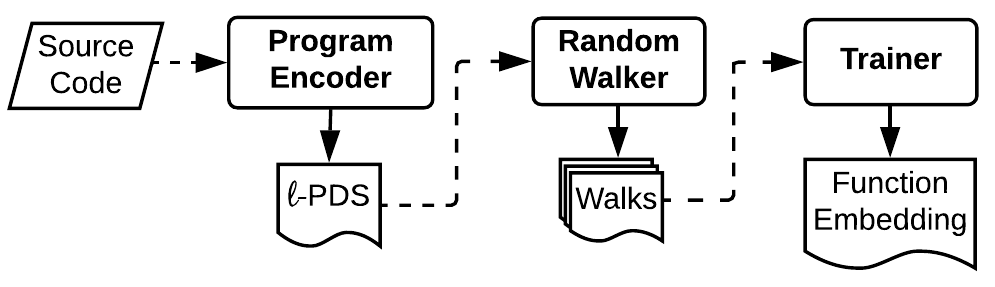}
\caption{\programtovec Architecture}
\label{fig:architecture}
\vspace{-1ex}
\end{figure}

\begin{figure*}
\begin{subfigure}[b]{.33\textwidth}
\begin{lstlisting}
void pci_disable_device(struct
		 pci_dev *dev) {
  
  struct pci_devres *dr = ...;
  if (dr == NULL) {
    dr->enabled = 0;
  }
  do_pci_disable_device(dev);
}

int snd_atiixp_create(...,
    struct pci_dev *dev,...) {
  
  struct atiixp *chip = ...; /*@\label{re:struct1}@*/
  if (chip == NULL) { /*@\label{re:eq}@*/
    pci_disable_device(dev); /*@\label{re:call1}@*/
    return -ENOMEM; /*@\label{re:error-code}@*/
  }
  int err =...;
  if (err < 0){
    pci_disable_device(dev); /*@\label{re:call2}@*/
    kfree(chip);
    return err;
  }
  chip->addr = ...; /*@\label{re:struct2}@*/
  ...
}
\end{lstlisting}
\caption{Simplified code from \cref{fig:spec-atiixp}.}
\label{fig:simplified-code}
\end{subfigure}
\begin{subfigure}[b]{.33\textwidth}
\includegraphics[scale=.4]{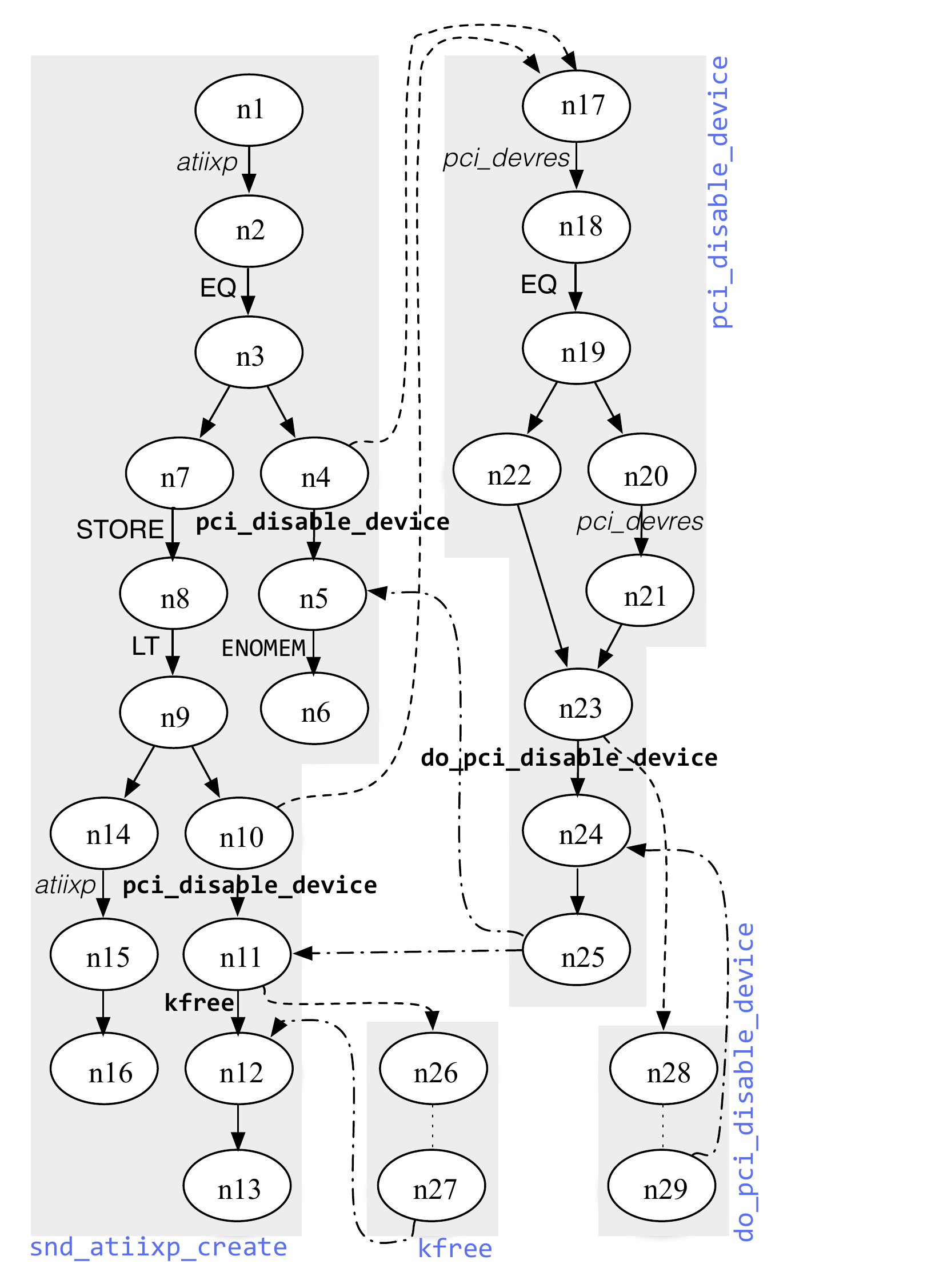}
\caption{Graphical representation of \lpds.}
\label{fig:icfg}
\end{subfigure}
\begin{subfigure}[b]{.33\textwidth}
\scriptsize
\begin{tabular}{l}
  {\small Rules for \texttt{snd_atiixp_create}:}\\
(1) $\intrarule{n_1}{n_2}$: \typelabel{atiixp}\\
(2)  $\intrarule{n_2}{n_3}$: \otherlabel{eq} \\
(3)  $\intrarule{n_3}{n_4}$ \\
(4)  $\intrarule{n_3}{n_7}$ \\
(5) $\callrule{n_4}{n_{17}}{n_5}$\\
(6) $\intrarule{n_4}{n_5}$: \functionlabel{pci_disable_device}\\
(7)  $\intrarule{n_5}{n_6}$: \errorlabel{enomem}\\
(8)  $\returnrule{n_6}$ \\
(9)  $\intrarule{n_7}{n_8}$: \otherlabel{store} \\
(10)  $\intrarule{n_8}{n_9}$: \otherlabel{lt} \\
(11)  $\intrarule{n_9}{n_{10}}$ \\
(12) $\intrarule{n_9}{n_{14}}$ \\
(13)  $\callrule{n_{10}}{n_{17}}{n_{11}}$ \\
(14) $\intrarule{n_{10}}{n_{11}}$: \functionlabel{pci_disable_device}\\
(15)  $\callrule{n_{11}}{n_{26}}{n_{12}}$ \\
(16) $\intrarule{n_{11}}{n_{12}}$: \functionlabel{kfree}\\
(17)  $\intrarule{n_{12}}{n_{13}}$ \\
(18)  $\returnrule{n_{13}}$ \\
(19)  $\intrarule{n_{14}}{n_{15}}$: \typelabel{atiixp} \\
(20)  $\intrarule{n_{15}}{n_{16}}$ \\
(21) $\returnrule{n_{16}}$\\
\\
  {\small Rules for \texttt{pci_disable_device}:}\\
(22) $\intrarule{n_{17}}{n_{18}}$: \typelabel{pci_devres} \\
(23) $\intrarule{n_{18}}{n_{19}}$: \otherlabel{eq} \\
(24) $\intrarule{n_{19}}{n_{20}}$ \\
(25) $\intrarule{n_{19}}{n_{22}}$ \\
(26) $\intrarule{n_{20}}{n_{21}}$: \typelabel{pci_devres} \\
(27) $\intrarule{n_{21}}{n_{23}}$ \\
(28) $\intrarule{n_{22}}{n_{23}}$ \\
(29) $\callrule{n_{23}}{n_{28}}{n_{24}}$ \\
(30) $\intrarule{n_{23}}{n_{24}}$: \functionlabel{do_pci_disable_device}\\
(31) $\intrarule{n_{24}}{n_{25}}$ \\
(32) $\returnrule{n_{25}}$ \\
\end{tabular}
\caption{\lpds rules.}
\label{fig:lpds}
\end{subfigure}
\caption{Running example.}
\label{fig:running-example}
\end{figure*}

The goal of \programtovec is to map a discrete set of functions to a
continuous vector space; that is, given a vocabulary $L$ of program
functions, each program function $\ell \in L$ is mapped to a
$d$-dimensional vector in $\mathbb{R}^d$. To accomplish this,
\programtovec generates a linearized representation of programs,
viz. ``sentences'' over a given vocabulary. \programtovec is the first
to use static program paths for this purpose. Intuitively, if we see
many program paths with a call to function \code{f2} after a call to
function \code{f1}, and paths with a call to \code{f3} after a call to
\code{f1}, then \code{f2} and \code{f3} should be embedded close to
each other.

A naive approach for linearizing a program is to generate a sentence
using the instructions along every valid interprocedural path in the
program. Such an approach has the following disadvantages: using the
entire instruction set would generate sentences with a very large
vocabulary; there are too many program paths for this approach to be
practical; and it does not capture the hierarchical structure of
programs.

The design of \programtovec addresses each of these disadvantages.
\programtovec abstracts each program instruction to reduce the
vocabulary of the sentences generated from the program path.  To
address the path explosion problem, \programtovec performs a random
walk of the program restricted to generate $\gamma$ paths of length at
most $k$ starting at a call to each function.
Lastly, on encountering a function call, the random walk either
outputs the function name itself, or decides to step into the function
definition.  This strategy of the random walk is able to capture the
hierarchical structure of programs: the context preceding the
function call can be linked to either the function call itself or to
the context in the body of the function being called.
\cref{fig:architecture} shows the three main components of
\programtovec.

\subsection{Program Encoder}
\label{sec:encoding}

We use a pushdown system (PDS) to model the set of valid
interprocedural paths in the program
\citep{DBLP:journals/scp/RepsSJM05}. A PDS is defined as
follows:

\begin{definition}
 \label{def:pds}
 A \emph{pushdown system} is a triple
 $\mathcal{P} = (P, \Gamma, \Delta)$ where $P$ and $\Gamma$ are finite
 sets, \emph{control locations} and \emph{stack alphabet},
 respectively. A \emph{configuration} of $\mathcal{P}$ is a pair
 $\langle p, w \rangle$, where $p \in P$ and $w \in \Gamma^*$.
 $\Delta$ contains a finite number of rules
 $\langle p , \gamma \rangle \hookrightarrow \langle p', w \rangle$,
 where $p, p' \in P$, $\gamma \in \Gamma$, and $w \in \Gamma^*$, which
 define a transition relation $\Rightarrow$ between configurations of
 $\mathcal{P}$ such that if
 $r = \langle p, \gamma \rangle \hookrightarrow \langle p', w
 \rangle$,
 then
 $\langle p, \gamma w' \rangle \Rightarrow \langle p', ww' \rangle$
 for all $w' \in \Gamma^*$.\qed
\end{definition}

\noindent We use $c \xRightarrow{r} c'$ to denote that the rule $r \in \Delta$
was used to transition from configuration $c$ to $c'$ of $\mathcal{P}$.
To model control flow of a program a single control location $p$, and
the following three types of rules
$r = \langle p, \gamma \rangle \hookrightarrow \langle p, w \rangle$
are sufficient: (i)~\emph{internal} rules with $|w| = 1$ that model
intraprocedural flow; (ii)~\emph{push} rules with $|w|=2$ that model
function calls, and (iii)~\emph{pop} rules with $|w|=0$ that model
function returns.

We use a mostly standard way of encoding an interprocedural
control-flow graph (ICFG) of a program as a PDS.  The main difference
is when a function call is encountered: given a call to function
\code{f} whose entry node is $\mathrm{e_f}$ on the ICFG edge
$\mathrm{n_1} \rightarrow \mathrm{n_2}$, we not only add the standard
call rule $\callrule{n_1}{e_f}{n_2}$, but also an internal rule
$\intrarule{n_1}{n_2}$. This new internal rule is akin to a summary
edge for the called procedure. As we will see, this internal rule
allows the random walk used by \programtovec (\cref{alg:random-walk})
to either step over or step into the function call.

A labeled PDS (\lpds) is a PDS in which each rule is associated with a
sequence of labels, and these labels are concatenated as the \lpds
makes its transitions. More formally:

\begin{definition}
\label{def:lpds}
  A \emph{labeled pushdown system} (\lpds) is a triple $\mathcal{L} =
  (\mathcal{P}, L, f)$, where $\mathcal{P} = (P, \Gamma, \Delta)$ is a
  PDS, $L$ is a finite set of labels, and $f: \Delta \rightarrow L^*$
  is a map that assigns a sequence of labels to each rule of
  $\mathcal{P}$. A \emph{configuration} of $\mathcal{L}$ is a pair
  $(c, l)$, where $c$ is a configuration of the PDS $\mathcal{P}$ and
  $l\in L^*$. $\Delta$ and $f$ define the transition relation
  $\Rightarrow_l$ between configurations of $\mathcal{L}$ such that if
  $c \xRightarrow{r} c'$, then $(c,l) \Rightarrow_l (c',l l')$, where
  $f(r) = l'$.  \qed
\end{definition}

\noindent We use $c \xRightarrow{r}_l c'$ to denote that the rule $r \in \Delta$
was used to transition from configuration $c$ to $c'$ of
$\mathcal{L}$.
In practice, we attach labels only to internal rules of the \lpds.  
We associate a unique label for each instruction category, error code,
\texttt{struct} type, and function.
Each instruction is mapped to a list of such labels as follows:
\begin{itemize}
  \item We classify instructions into categories such as LOAD, STORE,
    EQ, etc. 
   The internal rule associated with a particular instruction is labeled
   with the corresponding instruction category.
 \item Systems code defines specific constants that are used as error
   codes; see \cref{sec:ehnfer}. If such an error is used in the
   instruction, then we add the error-code label to the corresponding
   internal rule.
\item If the instruction loads or stores to a \texttt{struct}
  variable, then we add the \texttt{struct}-type label to the
  corresponding internal rule.
\item If the instruction is a function call, then we add the function
  label to the corresponding internal rule. 
\end{itemize}

\begin{example}
  \cref{fig:simplified-code} shows simplified functions
  \texttt{snd_create_atiixp} (from \cref{sec:motivating-example}), and
  \texttt{pci_disable_device}.  \cref{fig:icfg} shows a graphical
  representation of the corresponding \lpds. \cref{fig:lpds} lists the
  \lpds rules; we use the notation $r: l$ to mean that $f(r) = l$ in
  the \lpds. The instruction-category labels used in this example are
  \{\otherlabel{EQ}, \otherlabel{STORE}, \otherlabel{LT}\} (for
  simplicity, we do not include labels \otherlabel{LOAD} and
  \otherlabel{RET}); the \texttt{struct}-type labels are
  \{\typelabel{atiixp}, \typelabel{pci_devres}\}, the error-code
  labels are \{\errorlabel{ENOMEM}\}, and the function labels are
  \{\functionlabel{pci_disable_device}, \functionlabel{kfree},
  \functionlabel{do_pci_disable_device}\}.

  We describe the first 7 rules for the function
  \texttt{snd_atiixp_create} in \cref{fig:lpds}. The internal rule~(1)
  corresponds to \cref{re:struct1} in \cref{fig:simplified-code},
  where the variable \texttt{chip} of type \texttt{struct atiixp} is
  assigned. Thus, the rule is labeled with \texttt{struct}-type label
  \typelabel{atiixp}. The internal rule (2) corresponds to the
  equality expression on \cref{re:eq}, and is attached the instruction
  label \otherlabel{EQ}. Unlabeled rules~(3) and~(4) correspond to the
  \texttt{true} and \texttt{false} branches of the conditional on
  \cref{re:eq}. Call rule~(5) and internal rule~(6) are associated
  with the function call \texttt{pci_disable_device} on
  \cref{re:call1}. Note that the call rule is not labeled; the
  internal rule has a function label
  \functionlabel{pci_disable_device}. Finally, rule~(7) is given the
  error-code label \errorlabel{ENOMEM}, and corresponds to the return
  statement on \cref{re:error-code}.  \qef
 \end{example}

  \begin{algorithm}[t]
    \DontPrintSemicolon

    \KwIn{$\ell$-PDS $\mathcal{L} =(P, L, f)$, start label $\ell$, walk length $k$}
    \KwOut{walk = $\{\ell_1, \dots, \ell_n\}$}

    $\intrarule{n}{n'}:l \leftarrow \RandomKw(\{r:l | r \in \Delta
    \textrm{ and } \ell \in f(r) \})$\; \label{li:random-walk:random-rule}
    $c \leftarrow (\langle p, n'\rangle, l)$\; \label{li:random-walk:init-config}
    \For{$n\leftarrow 0$ \KwTo $k$}{
      $c \leftarrow \RandomKw(\{c' | c \xRightarrow{r}_l c' \textrm{
        for some } r \in \Delta \})$\; \label{li:random-walk:update-config}
    }
    \Return{$Labels(c)$}\;
    \caption{RandomWalk($\mathcal{L}$, $\ell$, $k$)}
    \label{alg:random-walk}
  \end{algorithm}

  \begin{algorithm}[t]
    \DontPrintSemicolon

    \SetKwFunction{Shuffle}{Shuffle}
    \SetKwFunction{RandomWalk}{RandomWalk}
    \SetKwFunction{SkipGram}{SkipGram}
    \KwIn{$\ell$-PDS $\mathcal{L} =(P, L, f)$, window size $w$, embedding size $d$,
      walks per label $\gamma$, walk length $k$}

    \KwOut{Vector representation for labels $\Phi : L \rightarrow \mathbb{R}^d$}

    $W \leftarrow \emptyset$\;
    \For{$i\leftarrow 0$ \KwTo $\gamma$} {
      \ForEach{$\ell_i \in L'$}{
        $W \leftarrow W \cup \RandomWalk(\mathcal{P}, \ell_{i}, k)$\;
      }
    }
    $\Phi \leftarrow \TrainModelKw(W, d, w)$\;\label{li:program2vec:train-model}
    \caption{\programtovec($\mathcal{L}$, $w$, $d$, $\gamma$, $k$)}
    \label{alg:program2vec}
  \end{algorithm}

\subsection{Random Walker}
\label{sec:random-walks}

\cref{alg:random-walk}
shows the algorithm to generate a random walk of a \lpds.
Given a set $S$, $\RandomKw(S)$ returns an element $s \in S$ that is
picked uniformly at random.  $Labels(\cdot)$ returns the sequence of
labels associated with a \lpds configuration.   Given a \lpds
$\mathcal{L}$, a start label $\ell$, and a walk length $k$, a random
walk is generated as follows. We randomly select a rule associated
with label $\ell$ (\cref{li:random-walk:random-rule}), and initialize
the configuration $c$ (\cref{li:random-walk:init-config}). Then, in
the loop at \cref{li:random-walk:update-config} we update the current
configuration $c$ by picking uniformly at random a next configuration
in the \lpds. Note that in \cref{def:lpds}, labels are concatenated
when the configuration is updated.

\begin{example}
  Consider $\ell = $ \typelabel{atiixp}, and $k = 10$. There are two
  rules associated with \typelabel{atiixp}: rules (1) and (19) from
  \cref{fig:lpds}. Assume we randomly pick rule (1). Thus, we start
  our random walk $w$ at rule (1) with label \typelabel{atiixp}. We
  then make 10 steps through the \lpds rules. Two possible
  random walks would be:\\
  $W_1 \eqdef$ \typelabel{atiixp} \otherlabel{EQ} \functionlabel{pci_disable_device} \errorlabel{ENOMEM}\\
  $W_2 \eqdef$ \typelabel{atiixp} \otherlabel{EQ} \typelabel{pci_devres}
  \otherlabel{EQ} \typelabel{pci_devres}
  \functionlabel{do_pci_disable_device}.

\noindent Note that during walk $W_1$ internal rule~(6) was chosen,
while $W_2$ descends into the call by choosing the call rule~(5).
\qef
\end{example}

\subsection{Model Trainer}
\label{sec:model-trainer}

Given a \lpds $\mathcal{L}$, a window $w$, a distance $d$, a number of
walks per label $\gamma$, and a walk length $k$, \programtovec
(\cref{alg:program2vec}) generates $\gamma$ walks for each label in
$L$, and uses them to train the model. The result is a vector
representation for labels $\Phi : L \rightarrow \mathbb{R}^d$.

$\TrainModelKw$ on \cref{li:program2vec:train-model} uses a neural network
to learn $\Phi$.
Traditional language models try to estimate the probability of seeing a label $\ell_i$ given the context
of the previous labels in the random walk; viz.\ $\textrm{Pr } \big(\ell_i | \ell_1,
\ell_2,\ldots, \ell_{i-1}\big)$.  Apart from learning the probability
distribution of label co-occurences, we also want to learn the
 embedding: $\Phi: L
\rightarrow \mathbb{R}^d$. Thus, our problem is to estimate
the likelihood:
$\textrm{Pr } \big(\ell_i | \Phi(\ell_1), \Phi(\ell_2),\ldots, \Phi(\ell_{i-1})\big)$.
\citet{word2vec} introduce a technique that uses a single-layer
fully-connected neural network to approximate this likelihood.
It uses a context of size $w$ both before and after the given word,
and considers the context as a set ignoring the ordering constraint.
This results in the the following optimization problem for computing
$\Phi$:
$\maximize_\Phi \textrm{log Pr }$ $\big(\ell_i | \{\Phi(\ell_{i-w}), \ldots, \Phi(\ell_{i-1}), \Phi(\ell_{i+1}), \ldots, \Phi(\ell_{i+w}) \} \big)$.
The implementation of \programtovec uses the implementation
of \citet{word2vec} provided in Gensim~\citep{rehurek-lrec}.

\subsection{Applications}

\begin{figure*}[!t]
    \centering
    \begin{subfigure}[t]{.33\textwidth}
        \centering
        \includegraphics[scale=0.48]{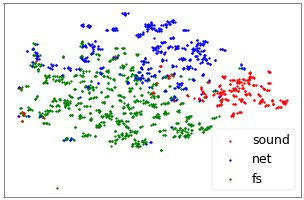}
        \caption{Three Linux Subsystems}
        \label{fig:components-tsne}
    \end{subfigure}
    \begin{subfigure}[t]{.33\textwidth}
        \centering
        \includegraphics[scale=0.48]{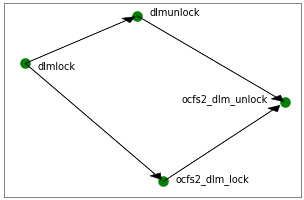}
        \caption{DLM Analogy}
        \label{fig:dlm-analogy}
    \end{subfigure}
    \begin{subfigure}[t]{.33\textwidth}
        \centering
        \includegraphics[scale=0.48]{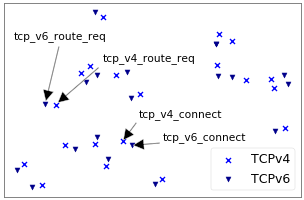}
        \caption{TCPv4 / TCPv6 Alignment}
        \label{fig:tcpv4-tcpv6-tsne}
    \end{subfigure}
    \caption{Visualization of \programtovec Applications}
    \label{fig:tsne}
\end{figure*}

The function embedding computed by \programtovec can be used in a variety
of applications:

\subsubsubsection{Identifying Function Synonyms.}
Function synonyms are close together in the \programtovec embedding, forming clusters by role.
\Cref{fig:tsne-intro} shows such clusters of function synonyms in the PCI sound drivers.
The K-means clustering algorithm is used to partition the word vectors
learned by \programtovec into function synonyms. 
The effectiveness of \programtovec to identify function synonyms in
the Linux kernel is evaluated in \cref{sec:programtovec-eval}.
Furthermore, the use of such function synonyms in mining
error-handling specifications is described in \cref{sec:ehnfer} and 
evaluated in \cref{sec:ehnfer-eval}.

\subsubsubsection{Subsystem Identification.}
Functions within subsystems
tend to be embedded closer to each other than functions between subsystems.
\cref{fig:components-tsne} shows a t-SNE projection of functions in three major
subsystems: sound, networking, and file systems.  File
systems such as GFS2 that rely on networking are closer to the networking
component than local-only file systems.

\subsubsubsection{Analogical Reasoning.}  The relationship between
OCFS2 Distributed Lock Manager (DLM) locking and unlocking is captured
by the analogy \texttt{dlmlock} : \texttt{dlmunlock} ::
\texttt{ocfs2\_dlm\_lock} : ?.  \Cref{fig:dlm-analogy} shows a PCA
plot of four functions belonging to the DLM.  Similar analogies can be
answered for other DLM locking and unlocking pairs, such as
\texttt{dlmlock\_remote} and \texttt{dlmunlock\_remote}.

\subsubsubsection{Alignment.}  The \programtovec embedding can be used
to match functions between related components.
\Cref{fig:tcpv4-tcpv6-tsne}
shows a t-SNE projection of TCPv4 and TCPv6 function pairs that have
been matched via Procrustes alignment~\citep{procrustes}.

\section{Func2vec Evaluation}
\label{sec:programtovec-eval}

We evaluated \programtovec against a runnable Linux kernel with all
file systems and PCI sound drivers included, roughly 2 million
LOC. The resulting \lpds consists of 5,407,483 stack locations
(nodes), 6,083,632 PDS rules, and 77,194 labels. For this experiment
the number of walks generated per label $\gamma$ was $100$, the walk
length $k$ was $100$, the vector dimension $d$ was $300$, and the
window size $w$ was $1$. To process the Linux kernel, \programtovec
requires approximately 24G of memory and two hours of compute time on
Amazon EC2 R4 instances. These instances use Intel Xeon E5-2686 v4
(Broadwell) Processors and DDR4 Memory.

\subsubsubsection{Gold Standard.} We chose the Linux kernel to
evaluate \programtovec because it is a prominent and important piece of
software. There does not, however, exist for Linux a benchmark of function
similarity that can be used as ground truth. Moreover, heuristics such as
relying on the naming conventions of functions or the natural language
interpretation of function names fall flat.

Function synonyms in Linux often follow a naming convention, such as
\texttt{snd_via82xx_free} and \texttt{snd_cmipci_free} for the \texttt{via82xx}
and \texttt{cmipci} sound drivers. Be that as it may, function synonyms do have
different names; \texttt{acpi_video_get_brightness} and
\texttt{intel_panel_get_backlight} each return the brightness level of the
backlight. Conversely, functions with similar names are not necessarily
synonyms; \texttt{rcu_seq_start} adjusts the current sequence number, while
\texttt{kprobe_seq_start} merely returns the current sequence number. Other
functions like \texttt{k8_mc0_mce} are cryptically named, without a single
natural language word.

Lacking an already existing benchmark, we created by hand a list of
\goldensetrelations\ relations between \goldensetfunctions\ unique functions to
serve as our gold standard. Of the \goldensetrelations\ relations,
\goldensetmust\ are assertions that two functions must be related, and
\goldensetmustnot\ assert that two functions must not be related.

\subsubsubsection{Evaluation Metrics.}  The \goldensetmust\ must
relations in our gold standard form \goldensetclasses\ equivalence
classes. For this evaluation, the must-not relations in the gold
standard are only used to check for consistency. We cluster the
\programtovec vectors with K-Means clustering and compare the
resulting clusters with the equivalence classes in the gold
standard. For each cluster $C_i$ and gold standard class $L_j$, the
precision, recall, and F-score are defined as follows.

\vspace{-1em}
\begin{center}
$$\text{Precision}(C_i, L_j) = \frac{|C_i \cap L_i|}{|C_i|}
\hspace{2em}
\text{Recall}(C_i, L_j) = \frac{|L_j \cap C_i|}{|L_j|}$$

$$F(C_i, L_j) = \frac{2 \times \text{Recall}(C_i, L_j) \times \text{Precision}(C_i, L_j)}{\text{Recall}(C_i, L_j) + \text{Precision}(C_i, L_j)}$$
\end{center}

To get an overall score that combines precision and recall, we use the
F-score over all gold standard classes
\citep[\S4.1]{amigo2009comparison}. Since an imperfect cluster may
partially overlap multiple gold standard classes we compute precision
and recall scores for the product of gold standard classes and K-Means
clusters. The precision and recall matrices are combined into a single
F-score matrix, penalizing a cluster for either including extra
functions or missing functions. The maximum F-score for each set of
synonyms in the gold standard is used, creating a mapping between
K-Means clusters and gold standard classes. The average F-score over
all classes in the gold standard is reported here, weighted by the
size of each gold standard class.

$$F = \sum_{i}{\frac{L_i}{N} \: \text{max}\{F(C_i, L_j)\}}$$

\subsubsubsection{Results.} Our evaluation of \programtovec\ shows
that it is capable of identifying relationships between functions in
the Linux kernel with both high precision and recall. We find that
\programtovec\ achieves an F-score of 0.77 out of 1.0 on our gold
standard of \goldensetmust\ relations between \goldensetfunctions\
unique functions.  \emph{This corresponds to a precision of
  \avgprecision\ and recall of \avgrecall\ over all classes in the
  gold standard.}


\section{Error-Handling Specifications}

\label{sec:ehnfer}
\begin{figure}
\includegraphics[scale=0.8]{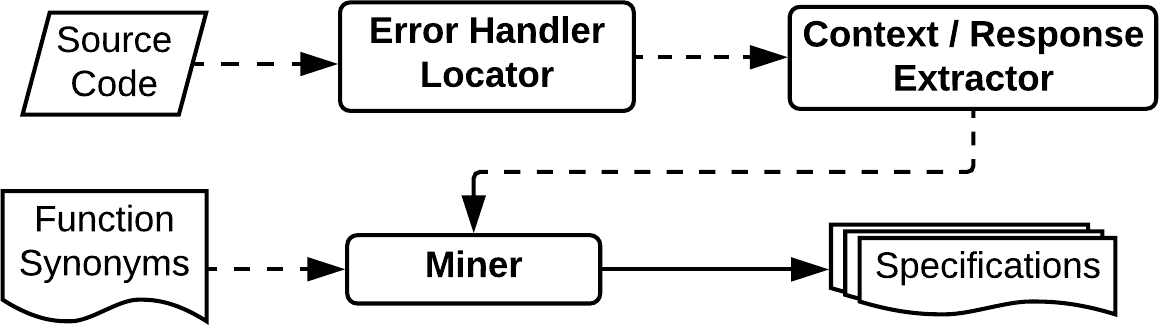}
\caption{Specification Mining Architecture}
\label{fig:mining-architecture}
\vspace{-4ex}
\end{figure}

To show the practical utility of identifying synonymous functions with
\programtovec, we present a detailed case study exploring their effect on mining
error-handling specifications. The major phases of the mining process are locating error
handlers, extracting error handler contexts and responses, and frequent itemset mining.
These phases are shown in \cref{fig:mining-architecture} and described in this
section.

\subsection{Defining Error-Handling Specifications}

We mine error-handling specifications based on the observation that the actions
performed after an error occurs (the error-handler response) frequently depend on the
actions carried out before the error occurred (the error-handler context). Such
a context and response pair define an error-handling specification. Our mining
data-set consists of all identified error-handler contexts and responses, where
each context and response is associated with an error handler that is uniquely
identified by the source location of a conditional branch.

\subsubsubsection{Error Handlers.}
An error handler is a piece of code that is executed
upon detection of an error. Our evaluation targets the Linux
kernel, which is written in C.  Without explicit error-handling
language constructs such as try/catch, locating error handlers in C
code must rely on some amount of domain knowledge.  We know that Linux
defines a specific set of integer error constants, referred to as
error codes. When returned from a function, these error codes denote
that an error has occurred. This error-handling mechanism is known as
the return-code idiom

\begin{example}
\label{ex:error-handlers}
The \code{snd_atiixp_create} function in \cref{fig:spec-atiixp}
contains the following five error handlers:
\begin{itemize}
\item \code{H1} (\cref{atiixp:eh1}) handles the error generated by
  \code{pci_enable_device} (\cref{atiixp:enable}).
  \item \code{H2} (\crefrange{atiixp:eh2-begin}{atiixp:eh2-end}) handles
  the error generated by \code{kzalloc}. 
  \item \code{H3} (\crefrange{atiixp:eh3-begin}{atiixp:eh3-end}) handles
  the error generated by \code{pci_request_regions}.
\item \code{H4} (\crefrange{atiixp:eh4-begin}{atiixp:eh4-end}) handles
  the error generated by \code{request_irq}.
\item \code{H5} (\crefrange{atiixp:eh5-begin}{atiixp:eh5-end}) handles
  the error generated by \code{snd_device_new}. \qef
\end{itemize}
\end{example}

We use two different techniques for locating error handlers in source code that
uses the return-code idiom: (1)~dataflow analysis based on
\citet{DBLP:conf/pldi/Rubio-GonzalezGLAA09} to locate conditional branches
testing values for error codes; and (2)~find basic blocks that return an error
code.

\begin{example}
  Consider the function \code{snd_atiixp_create} in
  \cref{fig:spec-atiixp}. The conditional statements for handlers
  \code{H1}, \code{H2}, \code{H3}, and \code{H5} at
  \cref{atiixp:enable,atiixp:regions,atiixp:irq,atiixp:device-new} all
  test the return value of a function, which the dataflow analysis
  reports may be an error code. These tests are either done through
  explicit assignment to the \texttt{err} variable, or directly as in
  the case of the statement on \cref{atiixp:irq}. \qef
\end{example}

\begin{example}
  Consider the handler \code{H4} on
  \cref{atiixp:ifkzalloc} in \cref{fig:spec-atiixp}. The conditional
  statement tests the return value of the function
  \texttt{kzalloc}. Because \texttt{kzalloc} returns \texttt{null}
  upon failure instead of an error code, the first technique does not
  detect this error handler. However, since an error is explicitly
  returned on the error path in \cref{atiixp:eh4-end}, the second
  technique identifies the handler. \qef
\end{example}

\subsubsubsection{Error-Handling Context and Response Sets.}
The context and response set of an error handler consist of the functions called
before and after an error is detected, respectively. In many cases, the order in
which these functions are called does not matter. Once the predicate associated
with the error handler is identified, the context and response sets are computed
via traversing the code paths in the backward and forward directions,
respectively.

\begin{example}
\label{ex:traces}
Context sets $C_H$ and response sets $R_H$ associated with the 
error handlers in \cref{ex:error-handlers}.\\
\small{
$C_{H1} = \emptyset$, $R_{H1} = \emptyset$ \\
$C_{H2} = $\{\code{pci_enable_device}\}, $R_{H2} = $\{\code{pci_disable_device}\} \\
$C_{H3} = $\{\code{pci_enable_device}, \code{kzalloc}\}, $R_{H3} = $\{\code{pci_disable_device}, \code{kfree}\} \\
$C_{H4} = $\{\code{pci_enable_device}, \code{kzalloc}, \code{pci_request_regions}\},\\
$R_{H4} = $\{\code{dev_err}, \code{snd_atiixp_free}\} \\
$C_{H5} =$\{\code{pci_enable_device},\code{kzalloc},\code{pci_request_regions},\code{request_irq}\},\\ 
$R_{H5} = $\{\code{snd_atiixp_free}\} \qef
}
\end{example}

\subsubsubsection{Error-Handling Specifications.}
An \emph{error-handling specification} is defined as an association
rule whose antecedent is the \emph{specification context} and
consequent is the \emph{specification response}.  This rule simply
means that the set of function calls in the context implies that the
set of function calls in the response are required to happen once an
error is detected.

\begin{definition}
\label{def:error-handling-specification}
An \emph{error-handling specification} $S$ is defined as $C_S \errorimplies R_S$,
where $C_S = \{c_1, c_2, \ldots, c_m\}$ is the context set of function calls for the specification $S$,
and $R_S = \{r_1, r_2, \ldots, r_m\}$ is the response set of function calls for the specification $S$.
\tinyqed
\end{definition}

\begin{example}
\label{ex:error-handling-specification}
The following error-handling specifications can be inferred for the atiixp sound driver:
\begin{itemize}
  \item $S_1 \eqdef$ \{\code{pci_enable_device}\} $\errorimplies$ \{\code{pci_disable_device}\}
  \item $S_2 \eqdef$ \{\code{pci_enable_device, kzalloc}\} $\errorimplies$ \{\code{kfree, pci_disable_device}\}
  \item $S_3 \eqdef$ \{\code{pci_enable_device, kzalloc, \\ pci_request_regions}\} $\errorimplies$ \{\code{snd_atiixp_free}\} \qef
\end{itemize}
\end{example}

\begin{definition}
\label{def:applicable-specification}
An error-handling specification $S \eqdef C_S \errorimplies R_S$ \emph{is applicable to} an error-handler $H$,
denoted by $S\appliesto H$,
iff $C_S \subseteq C_H$ and  $C_S \cup R_S \not\subseteq C_H$, 
where $C_H$ and $R_H$ are the context and response sets of the error handler $H$, respectively.
\tinyqed
\end{definition}

\begin{example}
\label{ex:applicable-specification}
Looking at the context and response sets for error-handlers \code{H1}
through \code{H5} in \cref{ex:traces} and the specifications in
\cref{ex:error-handling-specification}, we see that
$S_1 \appliesto H2$, $S_2 \appliesto H3$, $S_3 \appliesto H4$, and
$S_3 \appliesto H5$. \qef
\end{example}

The first term $C_S \subseteq C_H$ in \cref{def:applicable-specification} says that the
entire specification context must apply to the handler context. Otherwise the
specification does not speak to what the required response actions are.
The second term $C_S \cup R_S \not\subseteq C_H$ is added for
cases where the required response actions have already happened prior
to a particular error handler being reached, as illustrated by the following example.

\begin{figure}
\begin{lstlisting}[xleftmargin=1em,numbersep=1pt]
static struct inode *btrfs_new_inode(...) {
  // Most of the function is omitted
  path = btrfs_alloc_path();
  btrfs_free_path(path);   /*@\label{already-free}@*/
  ret = ...
  if (ret) { /*@\label{already-handler}@*/
    // btrfs_free_path not required
  }
}
\end{lstlisting}
  \vspace{-2ex}
\caption{Code snippet from the btrfs file system. The path is freed
  prior to the error handler on \cref{already-handler} is reached.}
\label{fig:alreadyhappened}
\vspace{-2ex}
\end{figure}

\begin{example}
Consider the specification $S \eqdef$ \{\code{btrfs_alloc_path}\}
$\errorimplies$ \{\code{btrfs_free_path}\}, and the code snippet from
the btrfs file system, shown in \cref{fig:alreadyhappened}:

The context set $C_H$ for the error handler $H$ at
\cref{already-handler} is \{\code{btrfs_alloc_path},
\code{btrfs_free_path}\}.  If we restrict
\cref{def:applicable-specification} to only contain the term
$C_S \subseteq C_H$, then we would say that specification $S$ is
applicable to the error handler $H$.  Clearly this is incorrect, as
the path has been allocated and then freed prior to the error handler
being reached.  Consequently, without the second term in
\cref{def:applicable-specification}, the error handler $H$ would be
flagged as a violation, even though the path has already been freed.
\qef
\end{example}

\begin{definition}
\label{def:satisfiable-specification}
An error-handling specification
$S \eqdef C_S \errorimplies R_S$ is \emph{satisfied by}
an error-handler $H$, denoted by $S\satisfiedby H$, iff $S\triangleright H$
and $R_S \subseteq R_H$, where $R_H$ is the response set of $H$.
\tinyqed
\end{definition}

Given a error-handling specification and an error handler, our miner
is able to report a violation of the specification using
\Cref{def:satisfiable-specification}.

\begin{example}
\label{ex:satisfiable-specification}
Looking at the context and response sets for the error-handlers in
\cref{ex:traces} and the specifications in
\cref{ex:error-handling-specification}, we see that
$S_1 \satisfiedby H2$, $S_2 \satisfiedby H3$, $S_3 \satisfiedby H4$,
and $S_3 \satisfiedby H5$. If the call on \cref{atiixp:eh5-begin} in
\cref{fig:spec-atiixp} was missing, then we would say that
$S_3 \not\satisfiedby H5$; i.e., $S_3$ would not be satisfied by
\code{H5}. \qef
\end{example}

The above definitions of error-handling specifications and
satisfiability can be extended to
handle \emph{synonymous functions}.
%
%
Let $F$ be the set of functions in the program we are mining.
$\Pi: F \rightarrow F$ is a said to be a \emph{partition function} iff
$\Pi(f_1) = \Pi(f_2)$ for all functions $f_1$ and $f_2$ that are
identical or synonyms.  We abuse notation slightly by extending the
partition function that applies to a single function to a set of
functions:
$\Pi(\{f_1, f_2,\ldots, f_n\}) = \{\Pi(f_1),
\Pi(f_2),\ldots,\Pi(f_n)\}$.
Similarly, given an error-handling specification
$S \eqdef C_S \errorimplies R_S$, we use $\Pi(S)$ to mean
$\Pi(C_S) \errorimplies \Pi(R_S)$.

\begin{definition}
\label{def:cross-implementation-error-handling-specification}
Given a partition function $\Pi$, 
a set $X$ of error-handling specifications is said to be a \emph{cross-implementation
error-handling specification} with respect to $\Pi$ iff
$\Pi(S) = \Pi(S')$ for all $S, S' \in X$.
\tinyqed
\end{definition}

\begin{example}
\label{ex:cross-implementation-error-handling-specification}
Let the partition function $\Pi$ be such that \\
$\Pi(\code{snd_atiixp_free}) = \Pi(\code{snd_intel8x0_free})$. Then \\
$\{\{\code{pci_enable_device, pci_request_regions}\} \errorimplies
\{\code{snd_attixp-}$ $\code{_free}\},$
$\{\code{pci_enable_device, pci_request_regions}\} \errorimplies
\{\code{snd-}$ $\code{_intel8x0_free}\}\}$
is a cross-implementation error-handling specification with respect to
$\Pi$, as described in \cref{sec:motivating-example}.  \qef
\end{example}

Using this notation, \cref{def:applicable-specification,def:satisfiable-specification}
can be naturally extended to cross-implementation error-handling specifications.

\begin{definition}
\label{def:applicable-cross-implementation-specification}

A cross-implementation error-handling specification $X$ \emph{is applicable to} an error-handler $H$,
denoted by $X\appliesto H$,
iff
there exists $S \in X$ such that $S\appliesto H$.
\tinyqed
\end{definition}

\begin{definition}
\label{def:satisfiable-cross-implementation-specification}

A cross-implementation error-handling specification $X$ \emph{is satisfied by} an error-handler $H$,
denoted by $X\satisfiedby H$,
iff
there exists $S \in X$ such that $S\satisfiedby H$.
\tinyqed
\end{definition}

\subsection{Mining Error-Handling Specifications}
\label{sec:mining}

Given the set of error handlers with their respective context and
response sets, we use frequent itemset mining to infer likely
error-handling specifications. Prior to mining, the error-handling
contexts and responses are normalized using the partition function
$\Pi$ (\cref{sec:approach}). The mined specifications involve the
normalized functions. The final step is to expand these specifications
into a set of specifications by replacing each normalized function
with the set of functions that maps to it.

\subsubsubsection{Frequent Itemset Mining.}
Let $\mathcal{T}$ be a set of transactions, where each transaction $T \in \mathcal{T}$ is a set of items.
A frequent itemset mining
algorithm returns the sets of items that frequently co-occur in the same
transaction in $\mathcal{T}$.
\begin{definition}
\label{def:support}
The \emph{support} of a set of items $I$ given a set of transactions
$\mathcal{T}$ is defined as
\(\support(I) \eqdef \left|\{ T \in \mathcal{T} \,\middle|\,
  I\subseteq T\} \right|.\) \tinyqed
\end{definition}
Frequent itemset mining algorithms take the minimum
support as a parameter, and return all sets of items that have a
support greater than or equal to the minimum support.

\section{Specification Mining Evaluation}
\label{sec:ehnfer-eval}

The experiments in this section are designed to answer the following
question: What is the effect of \programtovec on the quality of
  mined specifications?

\begin{table}
  \small
  \caption{Number of error handlers.} 
  \label{tab:handlers}
  \begin{tabular}{rrrrrr}
    \toprule
    Implementation & \# Handlers &&& Implementation & \# Handlers \\
    \cmidrule{1-2} \cmidrule{5-6}  
    btrfs & 1884 &&& OCFS2& 2103\\
    ext2 & 47 &&& Shared (VFS) & 887\\ ext4 & 680 &&& 48 sound drivers &
    3173\\ GFS2 & 539 &&& \textbf{Total} & \textbf{9313}\\ \bottomrule
    \end{tabular}
\end{table}

In this evaluation we focus on mining error-handling specifications, which is
considered to be an exceptionally difficult task
\cite{DBLP:conf/dsn/SahaL0LM13}. We mine error-handling specifications in 5
Linux file systems (btrfs, ext2, ext4, GFS2, and OCFS2), and 48 drivers. We
chose Linux file systems and drivers because of the dire consequences of error
handling defects, but the mining approach is general and can be applied to other
parts of Linux, or to any C program that uses the return-code idiom for error
handling. Frequent itemset mining is used to infer specifications in this
evaluation, but function synonyms can be used to enhance a wide variety of
mining techniques.

As shown in \cref{fig:mining-architecture}, our miner takes as input
(1) error handler context and response sets, and (2) synonymous
function information. Our implementation relies extensively on the
LLVM compiler infrastructure~\citep{DBLP:conf/cgo/LattnerA04}. To
locate error handlers, we use a combination of an existing LLVM-based
error-propagation analysis~\citep{DBLP:conf/pldi/Rubio-GonzalezGLAA09}
and a custom clang plugin.  \Cref{tab:handlers} shows the total number
of error handlers found. Function synonyms are identified by
\programtovec. Finally, Eclat/LCM \citep{eclat} is used to compute
frequent itemsets.

\begin{table}[tb]
  \caption{Specification mining with and without function synonyms.
    We inspected the top 150 specifications for file systems,
    and the top 50 for sound drivers.}
    \label{tab:spec-results} \centering
  \vspace{-2ex} 
  \small
  \begin{tabular}{llrrrrr}
    \toprule

    & Synonyms? & True & False & Min. Sup. & Avg. Precision \\
    \midrule
    File Systems & No & 125 & 25 & 31 & 0.77 \\
            & Yes & 141 & 9 & 76 & 0.92\\
    \midrule
    Drivers & No & 44 & 6 & 6 & 0.75 \\
            & Yes & 50 & 0 & 52 & 1.0 \\
    
    \bottomrule
  \end{tabular}
  \vspace{-2ex}
\end{table}

\subsection{Error-Handling Specifications}

We mined specifications for file systems and drivers separately. In each case
there were two runs: (1) without function synonyms, and (2) using function
synonyms. \cref{tab:spec-results} shows the results, which are described below. In
each case, we ranked the specifications by number of supporting examples and
inspected the highest ranked specifications. We measured the impact of function
synonyms by counting the number of true specifications in the top 150 for file
systems, and the top 50 for sound drivers. In addition to simply counting the
number of true and false positives, we also calculated the average precision,
which takes into account the relative position of true and false results
\cite{zhu2004recall}.

\subsubsubsection{File Systems.} Without synonyms, we found that 125 out of the
150 top specifications were true, and the remaining 25 were false positives.
With synonyms, 141 out of 150 top were true. Of the top 150 specifications, 130
rely on function synonyms. This means that only 20 specifications had a high enough
support without using synonyms to be in the top 150 (when using synonyms during
mining, not all specifications need make use of a synonym). Not only does the use of
function synonyms yield more true specifications in the top 150, but the true
specifications are more likely to appear higher in the list. This is reflected
in the higher average precision score.

\subsubsubsection{Drivers.} We inspected the top 50 specifications. Without
synonyms, 44 out of 50 top specifications are true specifications, and 6 are
false positives. With synonyms, all 50 specifications are true; there are no
false positives in the top 50. Of these 50, 46 made use of at least one pair
of function synonyms. As with the file systems, we see that the
quality of the mined specifications has improved, reducing the number of false
positives making visible specifications that would otherwise be entirely unreported.
This impact is also captured by the minimum support of any specification in the top 50,
i.e. the support of the 50th specification in the ranked list. Without using synonyms,
the 50th specification had a support of merely 6, in contrast to a support
of 52 when using synonyms.

\subsubsubsection{Examples.} \cref{tab:spec-examples} shows two examples of
cross-implementation error-handling specifications. The first example shows a
specification found across device drivers. All 14 functions found in the
response sets (e.g., \code{snd_korg1212_free}, \code{snd_intel8x0_free}, etc.)
are function synonyms reported by \programtovec. Being aware of these synonyms
helps the miner to determine that these specifications can be merged. This
results in a cross-implementation specification with a support of 57 and a rank
of \#4. Without synonyms, most of these individual specifications would not be
reported at all, given a reasonable support threshold of 5. The second example
is pulled from the GFS2 file system. In this case the miner and \programtovec,
working together, have identified the fact that the function
\code{gfs2_glock_dq_uninit} is synonymous with the actions performed by calling both
\code{gfs2_glock_dq} and \code{gfs2_holder_uninit}.


\begin{table}[t]
  \centering
  \small
  \caption{Examples of cross-implementation error-handling specifications.
    For brevity, we have removed the \code{pci_} prefix in
    \code{pci_enable_device}, and \code{pci_request_regions}.}
  \label{tab:spec-examples}
  \begin{tabular}{ll}
    \toprule
    Supp. & Specification \\
    \midrule
     11 & \{\code{enable_device}, \code{request_regions}\} $\errorimplies$ \{\code{snd_korg1212_free}\}  \\
     7 & \{\code{enable_device}, \code{request_regions}\} $\errorimplies$ \{\code{snd_intel8x0_free}\}  \\
     6 & \{\code{enable_device}, \code{request_regions}\} $\errorimplies$ \{\code{snd_intel8x0m_free}\}  \\
     5 & \{\code{enable_device}, \code{request_regions}\} $\errorimplies$ \{\code{snd_cmipci_free}\}  \\
     4 & \{\code{enable_device}, \code{request_regions}\} $\errorimplies$ \{\code{snd_ice1712_free}\}  \\
     4 & \{\code{enable_device}, \code{request_regions}\} $\errorimplies$ \{\code{snd_vt1724_free}\}  \\
     3 & \{\code{enable_device}, \code{request_regions}, \code{pci_ioremap_bar}\} $\errorimplies$   \\
      & \{\code{snd_cs4281_free}\}  \\
     3 & \{\code{enable_device}, \code{request_regions}\} $\errorimplies$ \{\code{snd_via82xx_free}\}  \\
     3 & \{\code{enable_device}, \code{request_regions}\} $\errorimplies$ \{\code{snd_ensoniq_free}\}  \\
     3 & \{\code{enable_device}, \code{request_regions}, \code{pci_ioremap_bar}\} $\errorimplies$  \\
      & \{\code{snd_ad1889_free}\}  \\
     2 & \{\code{enable_device}, \code{request_regions}\} $\errorimplies$ \{\code{snd_m3_free}\}  \\
     2 & \{\code{enable_device}, \code{request_regions}, \code{pci_ioremap_bar}\} $\errorimplies$  \\
      & \{\code{snd_atiixp_free}\}  \\
     2 & \{\code{enable_device}, \code{request_regions}\} $\errorimplies$ \{\code{snd_sonicvibes_free}\}  \\
     2 & \{\code{enable_device}, \code{request_regions}\} $\errorimplies$ \{\code{snd_es1968_free}\}  \\

    \addlinespace
    \midrule

    72 & \{\code{gfs2_holder_init}\} $\errorimplies$ \{\code{gfs2_holder_uninit}\}  \\
    49 & \{\code{gfs2_holder_init}\} $\errorimplies$ \{\code{gfs2_holder_uninit}, \code{gfs2_glock_dq}\}  \\
    45 & \{\code{gfs2_glock_nq}\} $\errorimplies$ \{\code{gfs2_holder_uninit}\}  \\
    39 & \{\code{gfs2_holder_init}, \code{gfs2_glock_nq}\} $\errorimplies$ \{\code{gfs2_holder_uninit}\}  \\
    29 & \{\code{gfs2_holder_init}, \code{gfs2_glock_nq}\} $\errorimplies$  \\
     & \{\code{gfs2_holder_uninit}, \code{gfs2_glock_dq}\}  \\
    23 & \{\code{gfs2_holder_init}\} $\errorimplies$ \{\code{gfs2_glock_dq_uninit}\}  \\
    22 & \{\code{gfs2_holder_init}\} $\errorimplies$  \\
     &  \{\code{gfs2_glock_dq_uninit}, \code{gfs2_holder_uninit}, \code{gfs2_glock_dq}\}  \\

    \bottomrule
  \end{tabular}
  \vspace{-2ex}
\end{table}

\subsection{Specification Violations}
\label{sec:violations}

\begin{figure}
\begin{lstlisting}[xleftmargin=1em,numbersep=1pt]
static int gfs2_get_flags(...) {
  ...
  /*@\bh@*/gfs2_holder_init(ip->i_gl, ...);/*@\eh@*/
  error = gfs2_glock_nq(&gh); /*@\label{confirmed-glock-nq}@*/
  if (error)
    // missing call to gfs2_holder_uninit
    return error; /*@\label{confirmed-return}@*/
  ...
  gfs2_glock_dq(&gh);
  gfs2_holder_uninit(&gh);
  return error;
}
\end{lstlisting}
  \vspace{-2ex}
  \caption{Bug found in GFS2. The function should not exit on
    \cref{confirmed-return} without calling
    \texttt{gfs2_holder_uninit}.}
\label{fig:gfs2}
\vspace{-2ex}
\end{figure}

Our miner can also be used to find violations to the specifications reported.
The GFS2 specification shown in~\cref{tab:spec-examples} led to the discovery of
two previously unknown bugs in the GFS2 file system. The patch we submitted to
fix these bugs was accepted by Red Hat and merged into Linux version 4.7.

\Cref{fig:gfs2} shows one of the two bugs. The function first calls
\texttt{gfs2_holder_init}, which acquires a reference to a glock. The
function then attempts to enqueue this holder structure. On the normal
path where \texttt{gfs2_glock_nq} succeeds, there is no problem, as at
the end of the function \texttt{gfs2_glock_dq} and
\texttt{gsf2_holder_unininit} are called. If \texttt{gfs2_glock_nq}
fails on \cref{confirmed-glock-nq}, however,
\texttt{gfs2_holder_uninit} is never called even though
\texttt{gfs2_holder_init} completed successfully. As is common with
error handling bugs, only in rare circumstances will this problem be
encountered because it requires \texttt{gfs2_glock_nq} to fail. But
when the bug is triggered the consequences are severe, resulting in an
inaccurate reference count for the glock.

\section{Related Work}
\label{sec:related}

\subsubsubsection{Distributed Representations.}
Distributed representations have been extensively studied in natural
language processing and cognition~\citep{hinton1986distributed}.
Recent advances have resulted in scalable approaches to computing such
distributed representations (or vector embeddings) given a corpus of
sentences; for instance, \texttt{word2vec}~\citep{word2vec}, and
Glove~\citep{pennington2014glove}.
DeepWalk~\citep{perozzi2014deepwalk} computes vector embeddings of
\emph{nodes} in a graph. DeepWalk is similar to \programtovec in that
they both use random walks to generate a corpus of sentences.
However, DeepWalk generates walks consisting of \emph{nodes}, while
\programtovec generates walks consisting of \emph{labels} along
edges. \programtovec also abstracts the program code into a
\lpds. \citet{ye2016word} apply vector representations to information
retrieval in software engineering by using \texttt{word2vec} on
documentation associated with code.

\citet{DBLP:conf/icse/NguyenNPN17} recently computed distributed
representations of API functions using \texttt{word2vec}. They
generated sentences using the program AST, as opposed to
interprocedural paths, and used their technique to migrate API usages
from Java to C\#.

\subsubsubsection{Error-handling Specification Mining.}
One of the key developments in the error-handling specification mining
literature has been the use of normal paths to mine specifications for
error-handling paths. This line of thought was first mentioned in
\citep{DBLP:conf/tacas/WeimerN05}, and was subsequently used in
several other papers \citep{DBLP:conf/tacas/WeimerN05,DBLP:conf/fase/AcharyaX09,
  DBLP:conf/icse/ThummalapentaX09, DBLP:conf/tacas/GouesW09,
  DBLP:conf/dsn/SahaL0LM13}.

\citet{DBLP:conf/tacas/WeimerN05} find association rules of the form
$FC_a \Rightarrow FC_e$, where function call $FC_a$ should be followed
by call $FC_e$, and $FC_e$ is found at least once in
exception-handling code. Improving on
\citep{DBLP:conf/tacas/WeimerN05},
\citet{DBLP:conf/icse/ThummalapentaX09} mine \textit{conditional}
association rules of the form
\((FC_c^1...FC_c^n) \wedge FC_a \Rightarrow FC_e^1...FC_e^n\),
which denotes a sequence of function calls prior to the target
function \(FC_a\)
that throws an exception, and then a sequence of recovery function
calls. \citet{DBLP:conf/fase/AcharyaX09} mine error-handling
specifications from interprocedural traces. Cleanup functions are
identified from error traces, which are then used along with normal
traces to find specifications. \citet{DBLP:conf/tacas/GouesW09}
broaden this notion of trace reliability to include a number of other
features (e.g., execution frequency, cloning, code age, density,
etc.), and significantly improve the false positive rate reported in
\citep{DBLP:conf/tacas/WeimerN05}. Collectively, these approaches have
been successful at finding defects in error-handling code that shares
function calls with normal paths. But there exist functions that are only
called on error paths, and that are only meaningful to Linux.  Thus
the correct use of these functions cannot be deduced from normal paths
or outside programs, and they would be missed by the above
approaches.

Much of the work on error-handling specifications has focused on
languages with exception-handling support, such as Java or C++.
Several approaches \citep{DBLP:conf/oopsla/WeimerN04,
  DBLP:conf/issta/BuseW08a, DBLP:conf/icse/ThummalapentaX09,
  DBLP:conf/tacas/WeimerN05} find error-handling specifications in
Java programs using static analysis. \citet{DBLP:conf/issta/BuseW08a}
infer and characterize exception-causing conditions, which are then
used as documentation. \citet{DBLP:conf/oopsla/WeimerN04} use dataflow
analysis to locate resource management mistakes in error-handling code
and propose a language extension to improve reliability. A common
mistake found might be the failure to release resources or to clean up
properly along all paths. Identifying blocks of error-handling code in
these languages is comparatively easy, but as we have shown, it is
more challenging to distinguish between normal and error-handling
paths in Linux.

\subsubsubsection{Implementation Inconsistencies.}
\citet{DBLP:conf/sosp/EnglerCC01} use the notion of internal
consistency to find programming errors. One of their techniques for
finding related pieces of code relied on the idiomatic use of function
pointers to define multiple implementations of a single
interface. \citet{cross-checking} compare multiple file systems by
leveraging the VFS interface to identify implementations of the same
functionality. These are complementary to our work.

\section{Conclusion}
\label{sec:conclusions}

We introduced the notion of \emph{function synonyms}: functions that
play a similar role in code.  Synonymous functions might be
syntactically dissimilar and might not be semantically equivalent.  We
presented \programtovec, an algorithm that maps each function to a
vector in a vector space such that function synonyms are grouped
together. Specifically, \programtovec computes a function embedding by
training a neural network on sentences generated using random walks of
the interprocedural control-flow graph of the program. We showed the
effectiveness and scalability of \programtovec by using
\goldensetclasses\ known classes of synonymous functions, with a total
of \goldensetfunctions\ functions, in the Linux kernel.

We also showed how \programtovec can improve the quality of
error-handling specifications. A challenge in mining these
specifications is that because error-handling code is not as common as
normal-execution code, the support of these specifications is often
too low. Our experimental evaluation on 5 Linux file systems and 48
Linux device drivers shows that this challenge is overcome by using
\programtovec to identify function synonyms across implementations,
and then using this information to mine error-handling specifications
with higher support.

\section*{Acknowledgments}

This work was supported in part by NSF grant CCF-1464439, and Amazon Web
Services Cloud Credits for Research.

\bibliographystyle{ACM-Reference-Format}
\bibliography{references}

\end{document}